\documentclass[12pt]{article}
\usepackage{amssymb}

\usepackage{graphicx}
\usepackage{amsmath}

\begin{document}

\title{ON \ HIGHER-DIMENSIONAL \ DYNAMICS}
\author{Paul S. Wesson$^{{}}$ \\
%EndAName
Department of Physics, University of Waterloo,\\
Waterloo, Ontario \ N2L 3G1, Canada\\
}
\maketitle

\bigskip

\bigskip


\bigskip


\bigskip

  Addresses: \ Mail to above. \ Phone (519) 885-1211 Ext. 2939;
Fax: (519) 746-8115; Email: wesson@astro.uwaterloo.ca

\newpage

\section*{Abstract}

\qquad Technical results are presented on motion in $N\;(>4)D$ manifolds to
clarify the physics of Kaluza-Klein theory, brane theory and string theory.
The so-called canonical or warp metric in 5D effectively converts the
manifold from a coordinate space to a momentum space, resulting in a new
force (per unit mass) parallel to the 4D velocity. The form of this extra
force is actually independent of the form of the metric, but for an unbound
particle is tiny because it is set by the energy density of the vacuum or
cosmological constant. It can be related to a small change in the rest mass
of a particle, and can be evaluated in two convenient gauges relevant to
gravitational and quantum systems. In the quantum gauge, the extra force
leads to Heisenberg's relation between increments in the position and
momenta. If the 4D action is quantized then so is the higher-dimensional
part, implying that particle mass is quantized, though only at a level of $%
10^{-65}$ gm or less which is unobservably small. It is noted that massive
particles which move on timeline paths in 4D can move on null paths in 5D.
This agrees with the view from inflationary quantum field theory, that
particles acquire mass dynamically in 4D but are intrinsically massless.\ A
general prescription for dynamics is outlined, wherein particles move on
null paths in an $N\;(>4)D$ manifold which may be flat, but have masses set
by an embedded 4D manifold which is curved.

\section{\protect\underline{Introduction}}

\ \ \ The motion of a test particle in a higher-dimensional manifold is a
prime way to investigate extensions of 4D Einstein theory. In 5D
Kaluza-Klein theory, older studies concentrated on the case where the 4D
spacetime was independent of the extra coordinate [1-6]. This condition was
relaxed in newer work on 5D induced-matter theory, where the 5D field
equations for apparent vacuum are broken down into 4D ones with an
energy-momentum tensor derived from the extra dimension [7-12].\ Dynamical
effects in 4D, when the metric is allowed to depend on one or more extra
coordinates, have also become the subject of studies in string and membrane
theory [13-16]. The main result is the appearance in 4D of extra forces
[13,17,18]. These are expected to be small in gravitational problems, but
could be significant in particle physics, where a unification of the
interactions could be achieved via 10D superstrings or 11D supergravity
[19,20]. In view of current interest in the subject, some results will be
given aimed at clarifying higher-dimensional dynamics.

\section{\protect\underline{Geodesics in ND and 4D}}

\qquad Consider an N-dimensional Riemannian manifold with a metric tensor $%
g_{AB}$ that depends on coordinates $x^{C}$, with a line element $%
dS^{2}=g_{AB}dx^{A}dx^{B}$, through which a particle moves along a path
described by an affine parameter $\lambda $. \ It contains a 4D submanifold
with line element $ds^{2}=g_{\alpha \beta }\,dx^{\alpha }dx^{\beta }$. \
Here and below it is instructive to concentrate on the 5D case, when Latin
indices run 0-4 and Greek indices run 0-3.

In general relativity, the particle moves along a geodesic which minimizes $%
s $ via $\delta \left[ \int ds\right] =0$. \ For a particle with mass $m$, $%
\lambda =s$ is normally chosen, the geodesic is non-null, and the 4-momentum
is defined by $p^{\alpha }\equiv mu^{\alpha }$ where the 4-velocity is $%
u^{\alpha }\equiv dx^{\alpha }/ds$. \ For a massless particle (photon), $%
\lambda $ is often unspecified because the geodesic is null and can be
obtained directly from the metric. \ The 4-velocities are conventionally
normalized for non-null and null paths via $u^{\alpha }u_{\alpha }=1,\,0$
respectively. \ However, $S$ contains $s$, and the former defines geodesics
via $\delta \left[ \int dS\right] =0$.

A question which arises in the literature is whether the particle should
follow a geodesic in ND or in 4D (in brane theory, this is connected with
whether the particle is constrained to move on the brane or can wander
through the bulk). \ The answer is that it is most natural to assume that
the motion is geodesic in $S$, provided that the extra terms which then
appear in the geodesic in $s$ are compatible with observation (see below). \
In this regard, it should be recalled that even in 4D the acceleration of
the particle $d^{2}x^{\alpha }/ds^{2}$ is \underline{not} a
covariantly-defined vector. \ The appropriate quantity is the absolute or
covariant derivative, which defines the path via $Du^{\alpha }/ds\equiv
du^{\alpha }/ds\,+\,\Gamma _{\beta \gamma }^{\alpha }\,u^{\beta }u^{\gamma
}=0$, where $\Gamma _{\beta \gamma }^{\alpha }$ are the Christoffel symbols.
\ It is the latter which yield the forces on the particle, which are thereby
recognized as being inertial in origin, meaning that they arise from the
motion of the reference frame. \ In ND, the same philosophy should hold. \
We can use $\delta \left[ \int dS\right] =0$, and the absolute derivative or
the Lagrange equations to obtain the dynamics, but the latter will in
general contain terms which arise from the motion with respect to the larger
reference frame.

The only comment which needs to be added to this concerns the case of null
geodesics with $dS=0$. \ This has been considered by several workers
[1,11,21]. \ It should be recalled that 4D causility is defined by $%
ds^{2}\geqslant 0$, and does not constrain $dS^{2}$ [5]. \ There is no
impediment to assuming that particles with $m\neq 0$ move along paths with $%
dS^{2}=0$, when their motions can be described consistently by choosing $%
\lambda =s$ .

There is, however, another issue which relates to geodesics and requires
notice. \ Geodesics in general relativity really describe accelerations, not
forces or changes in momentum. \ The distinction is often unnecessary,
because the mass is constant. \ But even in Newtonian mechanics, a rocket
changes mass as its fuel burns, feeling a force along the direction of its
motion. \ And in inflationary quantum field theory, particles are
intrinsically massless, gaining mass by a dynamical mechanism involving the
Higgs field [22]. \ As noted occasionally in the literature [23], the
correct dynamics in situations where the mass changes, follows not from $%
\delta \left[ \int ds\right] =0$ but from $\delta \left[ \int mds\right] =0$%
. \ That is, dynamics is a theory not of 4-velocities but of 4-momenta. \ In
this regard, it should be noted that there is no \underline{contradiction}
between the normalization condition used in relativity for the 4-velocities $%
u^{\alpha }u_{\alpha }=1$, and the condition used in particle physics for
the 4-momenta $p^{\alpha }p_{\alpha }=m^{2}$. \ (The latter is often written
$E^{2}-p^{2}=m^{2}$ and effectively uses the energy and 3-momentum to define
the rest mass of a particle.) \ It is just that the conventional geodesic
does not give any information about the origin or variability of mass, a
problem which is of central importance in cosmology.

The subjects of the preceding paragraphs, while perhaps familiar, underlie
much of the recent work which has been done on higher-dimensional dynamics
[1-23]. \ While it is not exclusive, we wish to concentrate in what follows
on an approach which resolves most of the issues raised above [7-13]. \
Specifically, we wish to present new results on metrics of the so-called
canonical or warp type in 5D (the extension to $N>5$ is straightforward). \
This has line element
\begin{equation}
dS^{2}=\frac{\ell ^{2}}{L^{2}}\,g_{\alpha \beta }\left( x^{\gamma },\ell
\right) dx^{\alpha }dx^{\beta }\,-\,d\ell ^{2},  \tag{2.1}
\end{equation}
where $x^{4}=\ell $ is the extra coordinate and $L$ is a constant length. \
Certain things are already known about this metric, and certain others may
be deduced from the comments made above. \ It is convenient to list these
here. \ (a) Mathematically (2.1) is general, insofar as the 5 available
coordinate degrees of freedom have been used to set $g_{4\alpha
}=0,g_{44}=-1 $. \ Physically, this removes the potentials of
electromagnetic type and flattens the potential of scalar type. \ (b) The
metric (2.1) has been extensively used in the field equations, which in
terms of the Ricci tensor are $R_{AB}=0$, and many solutions are known [11].
\ These include solutions for the 1-body problem [24] and cosmology [25]
which have acceptable dynamics, and solutions with the opposite sign for $%
g_{44}$ which describe waves [26]. \ (c) When $\partial g_{\alpha \beta
}/\partial \ell =0$ in (2.1), the 15 field equations $R_{AB}=0$ contain as a
subset the 10 field equations of general relativity, which in terms of the
Einstein tensor are $G_{\alpha \beta }=3g_{\alpha \beta }/L^{2}$. \ The
scale $L$ is thereby identified, in cosmology in terms of the cosmological
constant via $\Lambda =3/L^{2}$ and in other situations as the
characteristic size of the 4-space [10]. \ (d) This kind of local embedding
of a 4D Riemann space in a 5D Ricci-flat space can be applied to any N, and
is guaranteed by Campbell's theorem [27-30]. \ (e) The factorization in
(2.1) says in effect that the 4D part of the 5D interval is $\left( \ell
/L\right) ds$, which defines a \underline{momentum} space rather than a
\underline{coordinate} space if $\ell $ is related to $m$. \ This resolves
the issue of forces versus accelerations noted above. \ (f) Partial
confirmation of this comes from a study of the 5D geodesic and a comparison
of the constants of the motion in 5D and 4D [11,32,33]. \ In the Minkowski
limit, the energy of a particle moving with velocity $v$ is $E=\ell \left(
1-v^{2}\right) ^{-1/2}$ in 5D, which agrees with the expression in 4D if $%
\ell =m$. \ (g) The 5 components of the geodesic equation for (2.1) split
naturally into 4 spacetime components and an extra component. \ For $%
\partial g_{\alpha \beta }/\partial \ell \neq 0$, the former contain terms
parallel to the 4-velocity $u^{\alpha }$ as noted above for the case of a
rocket [10,33]. \ For $\partial g_{\alpha \beta }/\partial \ell =0$, the
motion is not only geodesic in 5D but geodesic in 4D.

\section{\protect\underline{The Nature of the Canonical Metric}}

\qquad The metric (2.1) is not mathematically unique but is physically rich,
which prompts a deeper examination of its nature.

Consider a 4D space with $ds^{2}=g_{\alpha \beta }dx^{\alpha }dx^{\beta }$
embedded in a 5D space with line element
\begin{equation}
dS^{2}=ds^{2}\,-\,d\ell ^{2}\;\;\;\;\;\;.  \tag{3.1}
\end{equation}
Then the transformation
\begin{equation*}
s\;\;\longrightarrow \;\;\ell \;\sinh \;\left( s/L\right)
\end{equation*}
\begin{equation}
\ell \;\;\longrightarrow \;\;\ell \;\cosh \;\left( s/L\right)  \tag{3.2}
\end{equation}
causes (2) to become
\begin{equation}
dS^{2}\,\;=\;\frac{\ell ^{2}}{L^{2}}\;ds^{2}\;-\;d\ell
^{2}\;\;\;\;\;\;\;\;\;\;.  \tag{3.3}
\end{equation}
This is of the canonical form, which is therefore recognized to be a
spherical form of a 2-plane (the opposite sign for $g_{44}$ may be obtained
by replacing the hyperbolic functions by their trigonometric counterparts).
\ Physically, there is an analogy with the angular momentum (per unit mass)
of a particle $rv$ moving with velocity $v$ at distance $r$ from the centre
of a circle. \ Recalling from above that $\ell $ plays the role of $m$ in
the canonical metric, we see that $mu^{\alpha }$ is the product of the
velocity in 4D with the distance in the orthogonal fifth direction. \ In
other words, $p^{\alpha }$ is a true 5D moment.

\qquad The above suggests that physically significant 4D structure may even
be present in a flat 5D manifold. \ The latter in spherical polars has line
element
\begin{equation}
dS^{2}=dt^{2}-dr^{2}-r^{2}d\Omega ^{2}-d\ell ^{2}\;\;\;\;\;\;\;\;\;,
\tag{3.4}
\end{equation}
where $d\Omega ^{2}\equiv \left( d\theta ^{2}+\sin ^{2}\theta d\phi
^{2}\right) $. \ Introducing a dimensionless parameter $\alpha $, consider
the transformation
\begin{equation*}
t\,\ \rightarrow \;\left( \frac{\alpha }{2}\right) t^{1/\alpha }\ell
^{1/(1-\alpha )}\left( 1+\frac{r^{2}}{\alpha ^{2}}\right) \;-\;\frac{\alpha
}{2(1-2\alpha )}\;\left[ t^{-1}\ell ^{\alpha /(1-\alpha )}\right]
^{(1-2\alpha )/\alpha }
\end{equation*}

\begin{equation*}
r\;\;\rightarrow \;\;rt^{1/\alpha }\;\ell ^{1/(1-\alpha
)\;\;}\;\;\;\,\;\;\;\;\;\;\;\;\;\;\;\;\;\;\;\;\;\;\;\;\;\;\;\;\;\;\;\;\;\;\;%
\;\;\;\;\;\;\;\;\;\;\;\;\;\;\;\;\;\;\;\;\;\;\;\;\;\;\;\;\;\;\;\;\;\;\;\;\;\;%
\;\;\;\;\;\;\;\;
\end{equation*}

\begin{equation}
\ell \;\;\rightarrow \;\;\left( \frac{\alpha }{2}\right) t^{1/\alpha }\ell
^{1/(1-\alpha )}\left( 1-\frac{r^{2}}{\alpha ^{2}}\right) \;+\;\frac{\alpha
}{2(1-2\alpha )}\;\left[ t^{-1}\ell ^{\alpha /(1-\alpha )}\right]
^{(1-2\alpha )/\alpha }\;\;\;\;\;  \tag{3.5}
\end{equation}
with $\theta \rightarrow \theta $, $\phi \rightarrow \phi $. \ Then some
algebra shows that (3.4) becomes
\begin{equation}
dS^{2}\,\;=\;\ell ^{2}dt^{2}\;-\;t^{2/\alpha }\ell ^{2/(1-\alpha )}\left(
dr^{2}\;+\;r^{2}d\Omega ^{2}\right) \;-\;\alpha ^{2}\left( 1-\alpha \right)
^{-2}t^{2}d\ell ^{2}\;.  \tag{3.6}
\end{equation}
This is the metric of a class of cosmological models first found as
solutions of the field equations $R_{AB}=0$ by Ponce de Leon [34]. \ They
are separable in $\ell ,\;t$ and reduce to the 4D Friedmann-Robertson-Walker
models with flat 3D sections on the hypersurfaces $\ell =$ constant. \ (The
dust or Einstein-de Sitter solution has $\alpha =3/2$ while the radiation
solution has $\alpha =2$.) \ For this reason they are often regarded as the
standard 5D cosmologies, but (3.5) shows that they are actually canonical
forms of 5D Minkowski space.

\qquad The same cannot be said of the standard 1-body solution of $R_{AB}=0$
[24]. \ This has the line element
\begin{equation*}
dS^{2}=\frac{\Lambda \ell ^{2}}{3}\left\{ \left[ 1-\frac{2M}{r}-\frac{%
\Lambda r^{2}}{3}\right] \;dt^{2}\;-\;\left[ 1-\frac{2M}{r}\;-\;\frac{%
\Lambda r^{2}}{3}\right] ^{-1}\;dr^{2}\right.
\end{equation*}

\begin{equation}
\left. -\;r^{2}\;d\Omega ^{2}\right\} -d\ell ^{2}\;\;\;\;\;,  \tag{3.7}
\end{equation}
where $M$ is a constant usually identified with the mass at the centre of
the 3-geometry. \ Metric (3.7) is \ pure canonical in the sense that it has
the form (2.1) with $\Lambda =3/L^{2}$ and $\partial g_{\alpha \beta
}/\partial \ell =0$. \ However, it is not 5D flat like (3.6), as may be
verified by computer. \ This agrees with the well-known fact that the 4D
Schwarzschild-de Sitter solution (given by the part inside the big brackets
in the last expression) can only be embedded in a flat space with dimension $%
N\geqslant 6$. \ However, since any 4D solution can be embedded in $%
N\geqslant 10$, the use of the canonical form obviously has relevance to
superstring theory.

\qquad The field equations $R_{AB}=0$ mentioned above clearly contain
physical information which is relevant to why the canonical metric (2.1) is
so effective. \ These 15 relations can in general be broken down into 1 wave
equation, 4 conservation equations, and 10 Einstein equations $G_{\alpha
\beta }=8\pi T_{\alpha \beta }$ with an effective or induced energy-momentum
tensor [35]. \ However, only the last quantity contains $u^{\alpha }$, and
via the 4D covariant derivative $T_{\;\;\;\;\;;\beta }^{\alpha \beta }=0$ is
usually interpreted as describing the dynamics of a fluid consisting of
particles. \ We will return to the field equations below, but let us here
take a look at particle dynamics.

\qquad This subject has been studied using both the geodesic equation [7,8]
and the Lagrange equations [10,12]. \ These approaches are compatible of
course, but the latter is the more instructive for investigating the special
status of (2.1). \ Thus consider a Lagrangian which generalizes (2.1) and
has the form
\begin{equation}
\mathcal{L}=\left\{ \frac{\ell ^{2}}{L^{2}}\;g_{\alpha \beta }\left(
x^{\gamma },\ell \right) \frac{dx^{\alpha }}{d\lambda }\;\frac{dx^{\beta }}{%
d\lambda }\;-\;\Phi ^{2}\left( x^{\gamma },\ell \right) \left( \frac{d\ell }{%
d\lambda }\right) ^{2}\right\} ^{1/2}\;\;\;\;.  \tag{3.8}
\end{equation}
This is dimensionless, and contains a scalar field $g_{44}=-\;\Phi ^{2}$
which is the classical analog of the Higgs potential that is responsible for
particle masses in quantum field theory [11,22]. \ The Lagrangian (3.8)
defines an action $I=\int \,\mathcal{L\,}\left( x^{A},\dot{x}^{A}\right)
\,d\lambda $, where $\lambda $ is an affine parameter along the path of the
particle and $\dot{x}^{A}\equiv dx^{A}/d\lambda $ is its 5-velocity. \ The
action is an extremum if
\begin{equation}
\frac{d}{d\lambda }\,\left( \frac{\partial \mathcal{L}}{\partial \dot{x}^{A}}%
\right) -\frac{\partial \mathcal{L}}{\partial x^{A}}=0\;\;\;\;\;\;\;\;\;\;\;%
\;\;\;\;\;.  \tag{3.9}
\end{equation}
The spacetime and extra components of this can be worked out explicitly once
$\lambda $ is chosen. \ A natural choice might appear to be $\lambda =S$
[8], in which case (3.9) is equivalent to
\begin{equation}
\frac{dU^{A}}{dS}\;+\;\Gamma _{BC}^{A}U^{B}U^{C}=0\;\;\;\;\;\;\;\;\;\;\;,
\tag{3.10}
\end{equation}
the 5D geodesic equation in $U^{A}\equiv dx^{A}/dS$ with appropriately
defined Christoffel symbols (see Section 2). \ However, this is undefined
for 5D null geodesics, and if another affine parameter $\lambda $ is used
instead then (3.10) acquires an extra term on the rhs equivalent to $U^{A}%
\mathcal{L}^{-1}d\mathcal{L}/d\lambda $ [36]. \ Also, the object of the
exercise is to understand 4D dynamics in $u^{\alpha }\equiv dx^{\alpha }/ds$
rather than 5D dynamics in $U^{A}\equiv dx^{A}/dS$. \ For these reasons, we
choose $\lambda =s$ [7,10,12]. \ We also choose $u_{\alpha }u^{\alpha }=1$
for a massive particle following a timelike 4D path. \ Then the substitution
of (3.8) into (3.9) results in two rather complicated expressions for the $%
\alpha ,4$ components of the motion, of which the second is the more
enlightening:
\begin{equation*}
\frac{\ell }{\Phi ^{2}}\left( \frac{\Phi ^{2}\dot{\ell}}{\ell }\right)
^{\cdot }\;-\;\frac{L^{2}\Phi \dot{\Phi}\dot{\ell}^{3}}{\ell ^{2}}\;-\;\frac{%
1}{L\Phi }\left( 1-\frac{L^{2}\Phi ^{2}\dot{\ell}^{2}}{\ell ^{2}}\right)
\left( 1+\frac{\ell u^{\alpha }u^{\beta }}{2}\;\;\frac{\partial g_{\alpha
\beta }}{\partial \ell }\;\;-\;\;\frac{L^{2}\Phi \dot{\Phi}\dot{\ell}^{2}}{%
\ell }\right)
\end{equation*}

\begin{equation}
\;\;\;\;\;\;\;\;\;\;\;\;\;\;\;\;\;\;\;\;\;\;\;\;\;\;\;\;\;\;\;\;\;\;\;\;\;\;%
\;\;\;\;\;\;\;\;\;=\;\;0\;\;\;\;\;\;\;\;\;\;.  \tag{3.11}
\end{equation}
Remarkably, this is satisfied with no constraint on the last parenthesis by $%
\left( L\Phi \dot{\ell}/\ell \right) ^{2}=1$, which by (3.8) implies $%
\mathcal{L}=0$. \ That is, the particle is travelling along a timelike path
in 4D but a null path in 5D.

\qquad We end this section by stating explicitly the equations of motion
which follow from (3.9) or (3.10). \ The spacetime components can be written
as a part which is geodesic in $s$ and an extra part:
\begin{equation}
\frac{du^{\mu }}{ds}\;+\;\Gamma _{\beta \gamma }^{\mu }\;u^{\beta }u^{\gamma
}=F^{\mu
}\;\;\;\;\;\;\;\;\;\;\;\;\;\;\;\;\;\;\;\;\;\;\;\;\;\;\;\;\;\;\;\;\;\;\;\;\;%
\;\;\;\;\;\;\;\;\;\;\ \   \tag{3.12}
\end{equation}

\begin{equation}
\;\;\;\;\;\;\;\;\;\;\;\;P^{\mu }\equiv \;\left( -g^{\mu \alpha }\;+\;\frac{%
u^{\mu }u^{\alpha }}{2}\right) u^{\beta }\;\frac{\partial g_{\alpha \beta }}{%
\partial \ell }\;\frac{d\ell }{ds}\;.  \tag{3.13}
\end{equation}
Here $F^{\mu }$ is a force per unit (inertial) mass, or acceleration. \ It
can be written as a sum of components normal and parallel to $u^{\mu }$, so $%
F^{\mu }=N^{\mu }\;+\;P^{\mu }$ where
\begin{equation}
N^{\mu }=\left( -g^{\mu \alpha }\;+\;u^{\mu }u^{\alpha }\right) u^{\beta }\;%
\frac{\partial g_{\alpha \beta }}{\partial \ell }\;\frac{d\ell }{ds}\;\;
\tag{3.14}
\end{equation}
\begin{equation}
P^{\mu }=\frac{-u^{\mu }}{2}\left( u^{\alpha }u^{\beta }\;\frac{\partial
g_{\alpha \beta }}{\partial \ell }\right) \;\frac{d\ell }{ds}%
\;\;\;\;\;\;\;\;\;.  \tag{3.15}
\end{equation}
The normal component obeys $N^{\mu }u_{\mu }=0$ (by construction), which is
the behaviour typical of Einstein gravity and Maxwell electromagnetism
[10,11]. \ The parallel component obeys

\begin{equation}
P^{\mu }u_{\mu }=\;\frac{-u^{\alpha }u^{\beta }}{2}\;\frac{\partial
g_{\alpha \beta }}{\partial \ell }\;\frac{d\ell }{ds}\;\equiv \;\beta
\;\;\;\;\;\;\;.  \tag{3.16}
\end{equation}
Here the 4-velocities are still normalized via $u^{\alpha }u_{\alpha }=1$
(see above and the next section). \ But the scalar quantity $\beta $ is
finite if $\partial g_{\alpha \beta }/\partial \ell \neq 0$ in the canonical
metric (2.1), and there is motion in the extra dimension as measured with
the particle's proper 4D time $s$. \ The quantity $\beta $ is a kind of
power per unit (inertial) mass. \ \underline{It has no analog in standard 4D
field theory}. \ The magnitude of $\beta $ depends on $d\ell /ds$, which is
given by the extra component of the equation of motion:
\begin{equation}
\frac{d^{2}\ell }{ds^{2}}\;-\;\frac{2}{\ell }\;\left( \frac{d\ell }{ds}%
\right) ^{2}\;+\;\frac{\ell }{L^{2}}\;=\;-\;\frac{1}{2}\;\left[ \frac{\ell
^{2}}{L^{2}}\;-\;\left( \frac{d\ell }{ds}\right) ^{2}\right] \;u^{\alpha
}u^{\beta }\;\frac{\partial g_{\alpha \beta }}{\partial \ell }\;\;.
\tag{3.17}
\end{equation}
This implies that there is no intrinsic state of rest for the particle in
the extra dimension. \ (Formally, the last equation is satisfied with $\ell
=\ell _{0}$, $u^{123}=0,\;u^{0}=1$ and $g_{\alpha \beta }=L^{2}/\ell
_{0}^{2} $, but then $d\ell =0$ and the metric reverts to a 4D one.) \ This
is basically because the 4D proper time has been used as a parameter for the
5D motion. \ Thus in general, $d\ell /ds\neq 0$ in (3.17), $\partial
g_{\alpha \beta }/\partial \ell \neq 0$ in (2.1) and $\beta \neq 0$ in
(3.16). \ The existence of finite scalar quantities like $\beta $ is
expected to be typical of the dynamics of any $N(>4)$-dimensional theory.

\section{\protect\underline{Extra Forces in $N(>4)D$ Theory}}

\qquad In section 2, it was noted that the word ``force'' has to be treated
with some caution, because theories like general relativity describe
accelerations while theories of particle physics describe momenta, and the
two concepts can only be consistently joined via a suitable definition of
mass. \ The canonical metric (2.1) opens a route to this, as well as
providing a number of other interesting results. \ In Section 3, it was seen
that the effectiveness of the canonical metric can be traced partly to the
fact that it is an algebraically convenient way to parametize a 5D manifold,
but mainly to the fact that it is a natural basis for 5D dynamics. \ In the
present section we wish to go beyond the canonical metric and note some
general results on forces.

\qquad It is straightforward to see that any $N(>4)D$ theory will
yield extra accelerations as viewed in 4D, which modulo an
appropriate definition of mass will be interpreted as extra
forces [13,17,18]. \ Given
an ND line element $dS^{2}=g_{AB}dx^{A}dx^{B}$, the N-velocities $%
u^{A}\equiv dx^{A}/dS$ are normalized for a non-null path via $U^{A}U_{A}=1$%
, and the path is extremized in terms of an ND covariant derivative via $%
U^{B}U_{\;\;\;\;;B}^{A}=0$. \ This when contracted gives $U_{A}F^{A}=0$,
where the $F^{A}$ are forces per unit (inertial) mass. \ However, this
implies $U_{\alpha }F^{\alpha }=-U_{(N-\alpha )}F^{(N-\alpha )}\neq 0$ when
viewed from 4D.

\qquad To relate this to what was done in Section 3, consider one extra
coordinate and the normalization condition
\begin{equation}
g_{\alpha \beta }\;(x^{\gamma },\ell )\;u^{\alpha }u^{\beta
}=1\,\;\;\;\;\;\;\;\;.  \tag{4.1}
\end{equation}
Differentiating this wrt an affine parameter $\lambda $ gives
\begin{equation}
g_{\alpha \beta ,\gamma }\;u^{\gamma }u^{\alpha }u^{\beta }\;+\;\frac{%
\partial g_{\alpha \beta }}{\partial \ell }\;\frac{d\ell }{d\lambda }%
\;u^{\alpha }u^{\beta }\;+\;2g_{\alpha \mu }\;\frac{du^{\mu }}{d\lambda }%
\;u^{\alpha }=0\;\;,  \tag{4.2}
\end{equation}
where $u^{\alpha }\equiv dx^{\alpha }/d\lambda $ and $g_{\alpha \beta
,\gamma }\equiv \partial g_{\alpha \beta }/\partial \ell $. \ Introducing
the Christoffel symbols and noting symmetries under the exchange of $\alpha $
and $\beta $, the first term on the lhs of (4.2) can be rewritten as

\begin{equation}
\left( g_{\alpha \beta ,\gamma }\;+\;g_{\alpha \gamma ,\beta }\;-\;g_{\beta
\gamma ,\alpha }\right) u^{\gamma }u^{\alpha }u^{\beta }=2g_{\alpha \mu
}\Gamma _{\beta \gamma }^{\mu }u^{\gamma }u^{\alpha }u^{\beta }\;\;\;\;\;.
\tag{4.3}
\end{equation}
Then (4.2) reads
\begin{equation}
2g_{\alpha \mu }u^{\alpha }\left( \frac{du^{\mu }}{d\lambda }\;+\;\Gamma
_{\beta \gamma }^{\mu }\;u^{\beta }u^{\gamma }\right) \;+\;\frac{\partial
g_{\alpha \beta }}{\partial \ell }\;\frac{d\ell }{d\lambda }\;u^{\alpha
}u^{\beta }=0\;\;\;.  \tag{4.4}
\end{equation}
With $F^{\mu }\equiv \left( du^{\mu }/d\lambda \;+\;\Gamma _{\beta \gamma
}^{\mu }u^{\beta }u^{\gamma }\right) $ as in (3.12) this says
\begin{equation}
u_{\mu }F^{\mu }=-\;\frac{u^{\alpha }u^{\beta }}{2}\;\frac{\partial
g_{\alpha \beta }}{\partial \ell }\;\frac{d\ell }{d\lambda }%
\;\;\;\;\;\;\;\;\;\;\;\;.  \tag{4.5}
\end{equation}
There is clearly a force per unit (inertial) mass parallel to $u^{\mu }$
given by

\begin{equation}
P^{\mu }=-\frac{u^{\mu }}{2}\;\left( u^{\alpha }u^{\beta }\;\frac{\partial
g_{\alpha \beta }}{\partial \ell }\right) \;\frac{d\ell }{d\lambda }%
\;\;\;\;\;\ .  \tag{4.6}
\end{equation}
When $\lambda =s$ this is identical to (3.15) above, which was derived
starting from the canonical metric (2.1). \ But here, we started from the
normalization condition (4.1) for a massive particle following a 4D timelike
path. \ This means that the existence of (4.6) does not depend on the form
of the metric. \ It is a consequence of defining and normalizing
4-velocities in the conventional way when spacetime is part of a bigger
manifold.

\qquad Given the manner in which 4D dynamics is conventionally set up, it is
difficult to conceive of any way in which a force parallel to the 4-velocity
could be interpreted other than by relating it to the mass of the particle
which feels it. \ For a particle moving under the influence of standard
gravity and the extra force, the equation of motion in spacetime is

\begin{equation}
\frac{du^{\mu }}{ds}\;+\;\Gamma _{\alpha \beta }^{\mu }\;u^{\alpha }u^{\beta
}=P^{\mu }\;\;\;\;\;,  \tag{4.7}
\end{equation}
with $P^{\mu }$ given by (4.6) with $\lambda =s$. \ Below, we will consider
some significant applications of (4.7). \ Here, as an illustration, we can
take a canonical metric with $g_{\alpha \beta }=(\ell ^{2}/L^{2})\eta
_{\alpha \beta }$ where $\eta _{\alpha \beta }$ = diagonal (1,-1,-1,-1). \
Then $\partial g_{\alpha \beta }/\partial \ell =(2/\ell )g_{\alpha \beta }$
and $u^{\alpha }u^{\beta }\partial g_{\alpha \beta }/\partial \ell =2/\ell $
in (4.6). \ The equation of motion (4.7) reads

\begin{equation}
\frac{du^{\mu }}{ds}\;=\;-\;\frac{u^{\mu }}{\ell }\;\frac{d\ell }{ds}\;\;,
\tag{4.8}
\end{equation}
which yields $\ell u^{\mu }=\ell _{0}$ where $\ell _{0}$ is a constant of
the (4D) motion. \ The last is clearly the momentum $mu^{\mu }$, confirming
that in canonical coordinates the extra coordinate plays the role of
particle mass. \ [See Section 2 and refs. 8, 10, 11 and 12. \ Equation (4.8)
above is the analog of what is sometimes called the rocket equation in
Newtonian mechanics, which just says that $d(m$v$)/dt=0$ or $d$v$/dt=-($v$%
/m)dm/dt$.] \ However, while $p^{\mu }=mu^{\mu }$ is a constant of the 4D
motion, it should be noted that $m=m(s)$ in general.

\qquad This cannot be fixed in the conventional approach to 4D dynamics,
except by appeal to some external condition. \ But in 5D dynamics it can,
notably by the extra component of the geodesic (3.17). \ This in general
requires a solution of the field equations, but as noted in Section 3 a
natural parametization of 5D geodesics is via $dS=0$. \ Then for the
canonical metric (2.1) we have

\begin{equation}
dS^{2}=0=\frac{\ell ^{2}}{L^{2}}\;ds^{2}-d\ell ^{2}\;\;\;\;,  \tag{4.9}
\end{equation}
which yields

\begin{equation}
\ell =\ell _{0}e^{\pm s/L}\;\;\;\;\;\;\;.  \tag{4.10}
\end{equation}
The rate of variation of $\ell $ depends on the characteristic dimension of
the 4-space $L$. \ As noted in Section 2, for pure-canonical metrics with $%
\partial g_{\alpha \beta }/\partial \ell =0$, $L=(3/\Lambda )^{1/2}$ where $%
\Lambda $ is the cosmological constant [10,11]. \ Thus from (4.10), with the
identification $\ell =m$, the rate of variation of the rest mass is given by
\begin{equation}
\frac{1}{m}\left| \frac{dm}{ds}\right| \simeq \;\frac{1}{m}\;\left| \frac{dm%
}{dt}\right| \;=\;\left( \frac{\Lambda }{3}\right) ^{1/2}\;\;.  \tag{4.11}
\end{equation}
The value of $\Lambda $ is severely constrained by astrophysical data
[37-39]. \ These indicate $\left| \dot{m}\right| /m\lesssim 2\;$x$\;10^{-18}$%
sec$^{-1}$ by (4.11), which is observationally acceptable. \ There are also
other constraints on extra forces like (4.6), but these are relatively weak
[40]. \ However, it should be noted that $\Lambda $ measures the energy
density of the vacuum in general relativity [11], and this could be larger
on small scales [22], so in principle mass variation and extra forces could
be measured.

\qquad To do this in practice, though, requires solutions of the
field equations. \ These in turn require the specification of a
system of coordinates or gauge. \ In this context, it should be
noted that the extra force $P^{\mu }$ of (4.6) for the 5D case is
a 4-vector. \ As such, it is covariant under the usual group of
4D coordinate transformations $x^{\mu }\rightarrow x^{\mu
}(x^{\nu })$, but will in general change under the group of 5D
coordinate transformations $x^{A}\rightarrow x^{A}(x^{B})$. \
This is inevitable given that the field equations $R_{AB}=0$ are
covariant in 5D, but may represent a problem as regards the
interpretation of observations made in 4D. \ This problem will be
greater in the (algebraically straightforward) extension of the
results of this section to $N(>5)D$. \ There is a considerable
relevant literature on gauges [11,19,20,41,42]. \ Fortunately, if
attention is restricted to dynamics there are only two natural
gauges, to which we now turn our attention.

\section{\protect\underline{The Einstein and Planck Gauges}}

\qquad It has been seen that the pure canonical gauge, namely (2.1) with a
factor $\ell ^{2}$ attached to an $\ell $-independent spacetime, is
remarkably successful as a basis for conventional 4D dynamics. \ This is
because the use of $\ell =m$ effectively converts the 4D part of the 5D
manifold from a coordinate space to a momentum space. \ However, that
success concerns the classical concept of momentum as the product of mass
and velocity. \ In modern quantum field theory the mass of a particle is not
defined \textit{a priori} [22], and even in old quantum theory the momentum
is described by a de Broglie wave and derived from a wave function. \ A
superior formulation of dynamics ought to address both the classical and
quantum nature of momentum. \ In this section, we will assume that the
differences in description are due to differences in gauge choices for a
higher-dimensional metric, and narrow the choices for the gauges using field
equations.

\qquad The latter are still the subject of discussion in ND field theory,
but in 5D there is a consensus that they are given by the Ricci tensor as

\begin{equation}
R_{AB}=0\;\;\;(A,B=0,123,4)\;\;\;\;\;\;\;.  \tag{5.1}
\end{equation}
Let us consider these for a generalized form of the pure canonical metric
(2.1) with line element
\begin{equation}
dS^{2}=\left( \frac{L}{\ell }\right) ^{2a}\;\bar{g}_{\alpha \beta
}(x^{\gamma })dx^{\alpha }dx^{\beta }-\left( \frac{L}{\ell }\right)
^{4b}d\ell ^{2}\;\;\;\;.  \tag{5.2}
\end{equation}
Here $a,b$ are constants which it is desired to constrain using (5.1). \
These 15, 5D relations can be decomposed into 4D ones under only the 4
coordinate conditions $g_{\alpha 4}=0$ [35]. \ For (5.2) it is convenient to
take the components in the order $AB=44,\;4\alpha ,\;\alpha \beta $. \ The
result is:

\begin{equation}
a^{2}-2ab+a=0\;\;\;\;\;\;\;\;\;\;\;\;\;\;\;\;\;\;\;\;\;\;\;\;\;\;\;\;\;\;\;%
\;\;\;\;\;\;\;\;\;\;\;\;\;\;\;\;\;\;\;\;\;\;\;\;\;\;\;\;\;\;\;\;\;\;
\tag{5.3}
\end{equation}

\begin{equation*}
V_{\alpha ;\beta }^{\beta
}=0\;\;\;\;\;\;\;\;\;\;\;\;\;\;\;\;\;\;\;\;\;\;\;\;\;\;\;\;\;\;\;\;\;\;\;\;%
\;\;\;\;\;\;\;\;\;\;\;\;\;\;\;\;\;\;\;\;\;
\end{equation*}

\begin{equation*}
V_{\alpha }^{\beta }\equiv \frac{1}{2\left| g_{44}\right| ^{1/2}}\left(
g^{\beta \sigma }\frac{\partial g_{\sigma \alpha }}{\partial \ell }%
\;-\;\delta _{\alpha }^{\beta }g^{\mu \nu }\frac{\partial g_{\mu \nu }}{%
\partial \ell }\right)
\end{equation*}
\begin{equation}
=\;\frac{3a\delta _{\alpha }^{\beta }}{\ell }\;\left( \frac{\ell }{L}\right)
^{2b}\;\;\;\;\;\;\;\;\;\;\;\;\;\;\;\;\;\;\;\;\;\;\;\;\;\;\;\;\;  \tag{5.4}
\end{equation}
\begin{equation}
G_{\alpha \beta }=\frac{(2a^{2}+2ab-a)}{\ell ^{2}}\;\left( \frac{L}{\ell }%
\right) ^{2a-4b}\bar{g}_{\alpha \beta }\;\;\;\;\;\;.  \tag{5.5}
\end{equation}
Here a semicolon denotes the usual 4D covariant derivative, $g_{\alpha \beta
}=(L/\ell )^{2a}\bar{g}_{\alpha \beta }$, $g_{44}=-(L/\ell )^{4b}$ and
indices have been raised and lowered using $g_{\alpha \beta }$ in order to
get the 4D Einstein tensor. \ This in mixed form is $G_{\alpha }^{\beta
}=(2a^{2}+2ab-a)\ell ^{-2}(\ell /L)^{4b}\delta _{\alpha }^{\beta }$, and as
usual $G_{\alpha }^{\beta }\equiv R_{\alpha }^{\beta }-R\delta _{\alpha
}^{\beta }/2$ in terms of the 4D Ricci tensor and Ricci scalar. \ The last
may be found by direct calculation, and is
\begin{equation}
R=\frac{-12a^{2}}{\ell ^{2}}\left( \frac{\ell }{L}\right) ^{4b}\;\;\;\;\;\;.
\tag{5.6}
\end{equation}
This determines the curvature of the 4D part of the manifold, which by (5.5)
has the form of a vacuum space with an effective cosmological constant $%
\Lambda =(2a^{2}+2ab-a)\ell ^{-2}(L/\ell )^{2a-4b}$. \ This is zero for $%
a=b=0$, in which case (5.2) describes general relativity embedded in a flat
and physically innocuous extra dimension. \ For $a=-1,b=0$ we have $\Lambda
=3/L^{2}$, and (5.2) is the pure canonical metric already discussed. \
Interestingly, the same value of $\Lambda $ results for $a=+1,\;b=+1$ which
implies that (5.5) would give the same 4D physics even though the metric
(5.2) does not have the canonical form. \ We will return to this below. \
Here we note that the 4D physics is contained in the 10 Einstein equations
(5.5), while the 4 conservation equations (5.4) are satisfied, and the 1
scalar relation (5.3) provides a constraint between the constants $a,b$. \
In this regard it should be noted that the relation $G=-R$ (which follows
from the definition of the Einstein tensor), when combined with (5.6)
reproduces (5.3), which is the only meaningful constraint.

\qquad The comments of the preceding paragraph imply that (5.2) as
constrained by (5.3) contains interesting physics in (5.5). \ For example,
the fact that the 4D cosmological constant depends in general on 5D
parameters opens the way to a resolution of the conflict in its size as
inferred from cosmology and particle physics [43-45]. \ However, it is
apparent that for dynamics there are two natural gauges, namely those with $%
a=-1,b=0$ and $a=+1,b=+1$. \ For these (5.2) has the forms

\begin{equation}
dS^{2}=\frac{\ell ^{2}}{L^{2}}\;\bar{g}_{\alpha \beta }(x^{\gamma
})dx^{\alpha }dx^{\beta }-d\ell ^{2}\;\;\;\;\;\;\;\;\;\;  \tag{5.7}
\end{equation}

\begin{equation}
dS^{2}=\frac{L^{2}}{\ell ^{2}}\;\bar{g}_{\alpha \beta }(x^{\gamma
})dx^{\alpha }dx^{\beta }-\frac{L^{4}}{\ell ^{4}}d\ell ^{2}\;\;\;\;\;.
\tag{5.8}
\end{equation}
Mathematically, these are equivalent since (5.7) $\rightarrow $ (5.8) under
the simple coordinate transformation
\begin{equation}
\ell \;\rightarrow \;L^{2}/\ell \;\;\;\;\;\;\;\;\;\;.  \tag{5.9}
\end{equation}
Physically, this corresponds to changing the way the rest mass of a particle
is described. \ We saw above that for the pure canonical metric the dynamics
implies the identification $\ell =m$. \ Let us restore physical units for
the speed of light $c$, the gravitational constant $G$ and Planck's constant
$h.$ \ Then (5.7) corresponds to the shift from gravitational ``units'' to
quantum ``units'', where the extra coordinate in (5.7), (5.8) is given
respectively by

\begin{equation}
\ell =\frac{Gm}{c^{2}}\;\;\;\;\;\;\;\;,\;\;\;\;\;\;\;\ell =\frac{h}{mc}%
\;\;\;\;\;\;\;\;\;.  \tag{5.10}
\end{equation}
There is nothing really fundamental about the presence of the dimensional
constants here. \ Dimensional analysis is an elementary group-theoretic
technique based on the Pi theorem [46-48]. \ The purpose of $c,G$ and $h$ in
(5.10) is merely to transpose the dimensions of mass to length so that it
can be geometrized. \ And since the dimensions of these quantities are $%
LT^{-1}$, $M^{-1}L^{3}T^{-2}$ and $ML^{2}T^{-1}$ and are not degenerate
[47], they can all be set to unity as is the common practice. \ This said,
it is convenient to rename the pure canonical metric (5.7) the Einstein
gauge and its other form (5.8) the Planck gauge.

\qquad The dynamics in these gauges can be studied for (5.7) by using
(3.12)-(3.17), and for both (5.7) and (5.8) by using (4.6), (4.7). \ As an
illustration, let us revisit the short calculation which led to (4.8) but
now in both gauges. \ That is, we take $g_{\alpha \beta }=(L/\ell )^{2a}\eta
_{\alpha \beta }$ so $\partial g_{\alpha \beta }/\partial \ell =-\left(
2a/\ell \right) g_{\alpha \beta }$ and $u^{\alpha }u^{\beta }\partial
g_{\alpha \beta }/\partial \ell =-2a/\ell $, which causes the 4D equation of
motion (4.7) with the parallel force (4.6) to read
\begin{equation}
\frac{du^{\mu }}{ds}=P^{\mu }=-\frac{u^{\mu }}{2}\left( u^{\alpha }u^{\beta
}\;\frac{\partial g_{\alpha \beta }}{\partial \ell }\right) \;\frac{d\ell }{%
ds}\;=\;\frac{au^{\mu }}{\ell }\;\frac{d\ell }{ds}\;\;\;\;.  \tag{5.11}
\end{equation}
This yields $u^{\mu }=\left( \ell /\ell _{0}\right) ^{a}$ where $\ell _{0}$
is a constant of the 4D motion. \ Alternatively, the last equation can be
written as

\begin{equation}
d\left( \ell ^{-a}u^{\mu }\right) =0\;\;\;\;\;\;\;\;.  \tag{5.12}
\end{equation}
This says that in the Einstein gauge $(a=-1)$ $\ell u^{\mu }$ is conserved,
while in the Planck gauge $(a=+1)\;u^{\mu }/\ell $ is conserved. \ That is,
by (5.10), the conserved quantities are the classical momentum $%
(G/c^{2})mu^{\mu }$ and the quantum momentum or inverse de Broglie
wavelength $(c/h)mu^{\mu }$. \ These are as expected, but as before in both
quantities $m=m(s)$ in general. \ If as in (4.9) we take a null 5D geodesic,
(5.2) gives
\begin{equation}
dS^{2}=0=\left( \frac{L}{\ell }\right) ^{2a}ds^{2}-\left( \frac{L}{\ell }%
\right) ^{4b}d\ell ^{2}\;\;\;.  \tag{5.13}
\end{equation}
This yields for \underline{both} gauges

\begin{equation}
\ell =\ell _{0}e^{\pm s/L}=\ell _{o}\exp \left[ \pm \left( \Lambda /3\right)
^{1/2}\;s\right] \;\;\;\;\;,  \tag{5.14}
\end{equation}
which is the same as (4.10) and has similar implications for mass variation.

\qquad This process is intrinsic to 5D dynamics and warrants a closer
examination, because it opens the way to understanding the Heisenberg
uncertainty relation. \ The latter is not a part of classical 4D dynamics,
and neither is the parallel 4D acceleration $P^{\mu }$ derived above. \ This
was calculated from the canonical metric (2.1) in (3.15) and has an
associated scalar power per unit (inertial) mass (3.16), but was also
calculated from the normalization condition (4.1) in (4.6) where it was
found to be of general form. \ This agrees with the\ Hamiltonian approach
to\ Kaluza-Klein theory, where a (4+1) split can always be performed in
order to recover Einstein theory [49-51]. \ In the Einstein and Planck
gauges, $P^{\mu }$ has the form (5.11), whose associated scalar is $P^{\mu
}u_{\mu }=(a/\ell )d\ell /ds$ where $a=\pm 1$. \ This quantity has no analog
in conventional classical dynamics, whose forces as was noted before obey $%
F^{\mu }u_{\mu }=0.$ \ However, an observer could interpret $P^{\mu }$ as
causing an anomalous change in momentum $d\bar{p}^{_{\mu }}$, such that by
(5.11)
\begin{equation}
\frac{1}{m}\;\frac{d\bar{p}^{\mu }}{ds}\;=\;P^{\mu }\;=\;\frac{au^{\mu }}{%
\ell }\;\frac{d\ell }{ds}\;\;\;\;.  \tag{5.15}
\end{equation}
This implies the associated scalar quantity

\begin{equation}
dx_{\mu }d\bar{p}^{\mu }=\frac{am}{\ell }\;d\ell \;ds\;\;\;\;\;\;\;.
\tag{5.16}
\end{equation}
For both the Einstein gauge and the Planck gauge, this reads

\begin{equation}
dx_{\mu }d\bar{p}^{\mu }=-dm\;ds\;\;\;\;\;\;\;\;\;.  \tag{5.17}
\end{equation}
That is, there is a Heinsenberg-type relation in 4D which depends on the
mass change associated with 5D. \ The relation (5.17) is general, insofar as
no use has been made of (5.14) for $m=m(s)$ which follows from (5.13) for a
null 5D geodesic. \ The generality of (5.17) may also be appreciated by
recalling from above that the momentum is really conserved along a 5D path,
so $d(mu^{\mu })=0$ or $mdu^{\mu }+u^{\mu }dm=0$, where the second term
represents an anomalous change in momentum $d\bar{p}^{\mu }=-u^{\mu }dm$, so
there is a scalar $dx_{\mu }d\bar{p}^{\mu }=-dx_{\mu }(dx^{\mu }/ds)dm=-dmds$%
.

\qquad Let us now ask how a 5D null path with $dS^{2}=0$ relates to the
motion of a particle of mass $m$ moving along a 4D timelike path with $%
ds^{2}>0$. \ We work in the Planck gauge with conventional units. \ Then
(5.14) gives $dm=\pm L^{-1}mds$, and (5.17) may be written
\begin{equation}
dx_{\mu }d\bar{p}^{\mu }=\frac{\left( mcds\right) ^{2}}{h}\;\left( \frac{%
\ell }{L}\right) \;\;\;\;\;\;\;.  \tag{5.18}
\end{equation}
The first term in parentheses here is the 4D action, which as usual is
defined and quantized for integers $n$ via

\begin{equation}
I\equiv \int mcds,\;\;\;\;\;\;\;\;\;\;\;dI=nh\;\;\;\;\;\;\;\;\;.  \tag{5.19}
\end{equation}
The second term in parentheses in (5.18) is the ratio of the Compton
wavelength of the particle $l=h/mc$ and the characteristic dimension of the
4-space it inhabits $L=(3/\Lambda )^{1/2}$. \ Now the 5D null condition
(5.13) in the Planck gauge $(a=1,\;b=1)$ shows that if the 4D action is
quantized via (5.19) then so must be the extra part of the 5D line element.
\ We therefore put $L/\ell =n$, which says that the Compton wavelength is an
excitation of the fundamental mode. \ The result is that (5.18) reads
\begin{equation}
dx_{\mu }d\bar{p}^{\mu }=nh\;\;\;\;\;\;\;\;,  \tag{5.20}
\end{equation}
which is Heisenberg's relation. \ The above approach can clearly be
generalized to other box sizes. \ But if free particles have masses set by
the energy density of the vacuum [22], constraints on the cosmological
constant [37-39] show that mass is quantized in units of
\begin{equation}
m=\frac{h}{cL}=\frac{h}{c}\left( \frac{\Lambda }{3}\right) ^{1/2}\;\lesssim
10^{-65}\;gm\;\;\;\;\;\;.  \tag{5.21}
\end{equation}
This is too small to detect with current methods, and miniscule compared to
the so-called Planck mass $(hc/G)^{1/2}\approx 10^{-5}gm$, which is seen to
be an artefact produced by a mixture of the Einstein and Planck gauges
discussed in this section.

\section{\protect\underline{Waves in $N(>4)D$ Theory}}

\qquad The preceding section showed that Heisenberg's relation
follows as a consequence of the extra force which results when a
causal 4D manifold is extended to a null 5D one. \ It is
therefore natural to ask if other aspects of particles, including
their wave nature, can be understood as manifestations of an
$N(>4)$-dimensional space. \ That quantum field theory and
general relativity are in principle compatible has been shown by
work on the Hartle-Hawking and Vilenkin wave functions [52-54]. \
And exact solutions of Einstein's equations are known which can
describe non-gravitational waves [26, 42, 55-60]. \ There is of
course the potential problem that the metric may be complex
[61-66], but this can be avoided if the view is taken that only
the physically-relevant quantities calculated from it need to be
real [26, 42, 60]. \ In this section we will therefore proceed to
see if it is possible to set up a consistent framework for 4D
wave mechanics in $N(>4)D$ theory, concentrating as before on the
5D case.

\qquad The 4D Klein-Gordon equation for a relativistic particle with zero
spin and finite mass should be derivable, based on what has been shown
above, from the 5D equation for a null geodesic. \ However, since the
Klein-Gordon equation is a second-order relation in a complex wave function,
we take the line element to have signature (+ - - - +). \ Then

\begin{equation}
dS^{2}=0=\left( \frac{L}{\ell }\right) ^{2a}ds^{2}+\left( \frac{L}{\ell }%
\right) ^{4b}d\ell ^{2}\;\;\;\;,  \tag{6.1}
\end{equation}
where $g_{\alpha \beta }=(L/\ell )^{2a}\bar{g}_{\alpha \beta }\;dx^{\alpha
}dx^{\beta }$ and $a=(-1,\;+1)\;$with $b=(0,\;+1)$ for the Einstein and
Planck gauges as before. \ The last relation is satisfied by
\begin{equation}
\ell =\ell _{0}e^{\pm is/L}\;\;\;\;\;\;,  \tag{6.2}
\end{equation}
which is the complex analog of (4.10). \ The mass (squared) involves $\ell
\ell ^{\ast }=\ell _{0}^{2}$ and is constant and real. \ Without loss of
generality we can take the upper sign in (6.2) and define a dimensionless
wave function
\begin{equation}
\psi =e^{is/L}\;\;\;\;\;\;\;\;.  \tag{6.3}
\end{equation}
This satisfies a hierachy of wave equations
\begin{equation}
\frac{d^{n}\psi }{ds^{n}}\;=\;\left( \frac{i}{L}\right) ^{n}\;\psi
\;\;\;\;\;\;\;\;\;\;,  \tag{6.4}
\end{equation}
where we are interested primarily in the cases $n=1,2$. \ [The $n$ in (6.4)
should not be confused with that in (5.19).] \ The first-order equation of
(6.4) implies $i\psi /L=dx/ds=(\partial \psi /\partial x^{\alpha
})(dx^{\alpha }/ds)=(\partial \psi /\partial x^{\alpha })u^{\alpha }$ or $%
1=(L/i\psi )(\partial \psi /\partial x^{\alpha })u^{\alpha }$. \ But $%
1=u^{\alpha }u_{\alpha }$, so
\begin{equation}
u_{\alpha }=\frac{L}{i\psi }\;\frac{\partial \psi }{\partial x^{\alpha }}%
\;\;\;\;.  \tag{6.5}
\end{equation}
This for the fundamental mode of the Planck gauge with $L=h/mc$ just says $%
p_{\alpha }=(h/ic\psi )\partial \psi /\partial x^{\alpha }$, which is the
usual prescription for obtaining the momenta from the wave function. \ The
second-order equation of (6.4) can be treated similarly, and with (6.5)
yields
\begin{equation}
\frac{u^{\alpha }u^{\beta }}{\psi }\;\;\frac{\partial ^{2}\psi }{\partial
x^{\alpha }\partial x^{\beta }}\;+\;\frac{1}{L^{2}}\;+\;\frac{iu_{\alpha }}{L%
}\;\frac{du^{\alpha }}{ds}\;=\;0\;\;\;\;\;\;\;.  \tag{6.6}
\end{equation}
The imaginary part of this is $u_{\alpha }du^{\alpha }/ds=0$ or the usual
orthogonality relation. \ The parallel acceleration (4.6) or (5.11) which
follows from the metric in form (5.2) does not appear, which agrees with the
fact that the effective mass is a constant for the metric in form (6.1). \
In this regard, it is instructive to consider the geodesic equation $dU^{\mu
}/d\lambda +\Gamma _{\beta \gamma }^{\mu }U^{\beta }U^{\gamma }=0$ with $%
U^{\alpha }\equiv dx^{\alpha }/d\lambda $, $d\lambda =e^{\mp is/L}ds$ and $%
\Gamma _{\beta \gamma }^{\mu }$ constructed from $g_{\alpha \beta }=(L/\ell
)^{\pm 1}\bar{g}_{\alpha \beta }$, as implied by (6.1). \ This geodesic, it
may be verified, splits naturally into real and imaginary parts:
\begin{equation}
\frac{du^{\mu }}{ds}\;+\;\bar{\Gamma}_{\beta \gamma }^{\mu }u^{\beta
}u^{\gamma }=0  \tag{6.7}
\end{equation}

\begin{equation}
u^{\mu }=\bar{g}^{\mu \alpha }\left( \bar{g}_{\alpha \gamma }\;\frac{%
\partial s}{\partial x^{\beta }}+\bar{g}_{\alpha \beta }\frac{\partial s}{%
\partial x^{\gamma }}\;-\;\bar{g}_{\gamma \beta }\;\frac{\partial s}{%
\partial x^{\alpha }}\right) \;u^{\beta }u^{\gamma }\;\;\;\;\;.  \tag{6.8}
\end{equation}
Here $u^{\mu }\equiv dx^{\mu }/ds$ and $\bar{\Gamma}_{\beta \gamma }^{\mu }$
is constructed from $\bar{g}_{\alpha \beta }$. \ We see from (6.7) that the
motion is geodesic in the embedded 4-space. \ And (6.8) is identically
satisfied if $s=\int u_{\alpha }dx^{\alpha }$ so $\partial s/\partial
x^{\alpha }=u_{\alpha }$ as usual for the 4-interval. \ In other words, the
4D dynamics is standard. \ Returning now to (6.6), its real part may be
rewritten by noting that $u^{\alpha }u^{\beta }\partial ^{2}\psi /\partial
x^{\alpha }\partial x^{\beta }=\bar{g}^{\alpha \beta }\partial ^{2}\psi
/\partial x^{\alpha }\partial x^{\beta }$, which can be shown using (6.5), $%
u_{\alpha }=\partial s/\partial x^{\alpha }$ and $u^{\alpha }u_{\alpha }=1$.
\ Then for the fundamental mode of the Planck gauge, the real part of (6.6)
reads
\begin{equation}
\bar{g}^{\alpha \beta }\;\frac{\partial ^{2}\psi }{\partial x^{\alpha
}\partial x^{\beta }}\;+\;\frac{m^{2}c^{2}\psi }{L^{2}}\;=\;0\;\;\;\;\;\;\;%
\;.  \tag{6.9}
\end{equation}
This is the standard 4D Klein-Gordon equation.

\qquad The preceding paragraph started with the complex metric (6.1) and
ended with the relativistic wave equation (6.9). \ Both are statements about
dynamics, and neither uses the field equations. \ Solutions of the latter of
relevant type were mentioned above [26, 42,55-60]. \ It would be
inappropriately long to discuss these here; but to show that there is a
match between the dynamics and the field equations, let us take the metric
(6.1) in the Planck gauge $(a=1,\;b=1)$ and consider the 5D field equations $%
R_{AB}=0\;(A,B=0,123,4)$. \ The latter may be shown to be satisfied, either
tardily using algebra [35] or quickly using a computer package [67], by

\begin{equation}
dS^{2}=\left( \frac{L}{\ell }\right) ^{2}\left( dt^{2}-e^{i(\omega
t+k_{x}x)}dx^{2}-e^{i(\omega t+k_{y}y)}dy^{2}-e^{i(\omega
t+k_{z}z)}dz^{2}\right) +\left( \frac{L}{\ell }\right) ^{4}d\ell .
\tag{6.10}
\end{equation}
Here $k_{xyz}$ are arbitrary wave-numbers along the Cartesian axes and $%
\omega $ is a frequency fixed by the field equations as $\omega =\pm 2/L$. \
This solution may be regarded as the canonical one in 5D, since not only
does the Ricci tensor $R_{AB}$ vanish but the Riemann tensor $R_{ABCD}$ does
also. \ That is, (6.10) describes a wave propogating in the 4D part of a
\underline{flat} 5D manifold. \ However, the 4D part of (6.10) is \underline{%
curved}. \ In the induced-matter picture [11], it is curved by a
cosmological constant $\Lambda =-3\ell ^{2}/L^{4}$. \ If the latter is
modelled as in general relativity by a classical pressure and density, the
wave in 4D is supported by a medium with the equation of state $8\pi p=-8\pi
\rho =-\Lambda =3\ell ^{2}/L^{4}$ typical of the classical vacuum. \ Another
way of appreciating what is involved here is by considering the extra
coordinate $\ell $, or equivalently the inertial rest mass of the particle $%
m $. \ By (6.2), $\ell $ oscillates in and out of the 4D spacetime
hypersurface defined by $s$. \ The average value of $\ell $ is zero,
agreeing with the fact that $R_{ABCD}$ describes a flat 5D manifold; but the
average value of its square is finite, agreeing with the fact that the 4D
manifold is curved. \ By (6.10) directly, or by (5.6) modulo a sign due to
the change in signature from (5.2) to (6.1), the 4D Ricci scalar is

\begin{equation}
R=\frac{12\ell ^{2}}{L^{4}}=\frac{12m^{2}c^{2}}{h^{2}}\;\;\;\;\;\;\;\;.
\tag{6.11}
\end{equation}
Here we have taken the fundamental mode in the Planck gauge ($\ell =L=h/mc$%
). \ The last relation just says that the scalar curvature of the 4D space
is set by the Compton wavelength or mass of the particle which inhabits it.
\ This agrees with\ Mach's principle [11], and with the idea from
inflationary quantum field theory that particles are intrinsically massless
[22]. \ There is no contradiction, as long as the view is taken that
particles with finite rest mass are 4D objects in a 5D vacuum.

\qquad The generalization of the above is expected to be straightforward,
both for spin$-1/2$ particles described by the Dirac equation in 5D [68,69]
and spin-0 particles described by scalar wave equations in ND [11,70]. \ In
the latter context, it is clear that the defining equation for classical
dynamics in terms of the N-velocities should be $U^{A}U_{A}=0$. \ The
corresponding quantum wave function can be derived from the metric as fixed
by solutions of $R_{AB}=0\;(A,B=0,123,...N)$. \ Probabilities should be
defined from the metric using the element of proper volume $\left( \left|
g\right| \right) ^{1/2}dx^{N}$, so $\left| g\right| $ will be a non-trivial
factor. \ [This is already evident in 5D from metrics like (6.10), where the
4D wavefunction may be augmented by $g_{44}$ to yield a 5D one whose extra
component will be related to the spectrum of particle masses.] \ It is
expected to report on these issues in the future.

\section{\protect\underline{Conclusion}}

\qquad In Section 2, we noted that the canonical 5D metric (2.1)
justifies its name by providing a basis for $N(>4)D$ geodesics
and leading to many useful results. \ The utility of the
canonical (or warp) metric can be understood as a result of
embeddings, which on 4D hypersurfaces reduces to physically
acceptable solutions for cosmology (3.6) and the 1-body problem
(3.7). \ However, as shown elsewhere in Section 3, its efficacy
is mainly due to the fact that it converts a
\underline{coordinate} space to a \underline{momentum} space,
\underline{with the extra} \underline{coordinate playing the role
of rest mass even for null 5D geodesics}. \ In general, the
velocity in the extra dimension results in a new inertial force
(per unit mass) given by (3.15). \ This and its associated power
(per unit mass) given by (3.16) have no analog in conventional 4D
dynamics. \ The same is true of any non-trivial ND manifold. \ In
Section 4, it was seen that the normalization condition (4.1)
results in an extra acceleration parallel to the 4-velocity which
has the form (4.6) independent of the coordinate system. \ This
can most logically be handled by connecting it to the change in
the (inertial) rest mass of the particle which feels it, which
means that the 4D motion is technically non-geodesic in the
4-velocity (4.7), even though it agrees with the conventional law
for the conservation of the 4-momentum as applied (say) to the
motion of a rocket (4.8). \ However, the rate of variation of
mass in the Minkowski limit is set by the cosmological constant
as in (4.11) and is tiny, so for apparently free particles there
is no conflict with observed dynamics. \ In Section 5, the field
equations (5.1) were used to constrain the form of a generalized
canonical metric (5.2), leading to the recognition of two natural
choices for the extra coordinate in (5.7) and (5.8). \ These are
related by an elementary coordinate transformation (5.9), whose
physical meaning is however significant: classical physics uses
the gravitational constant of Newton or Einstein to geometrize
the mass, while quantum physics uses Planck's constant to
geometrize the mass, as in (5.10). \ That is, classical and
quantum dynamics in 4D are descriptions of 5D dynamics in what
can be termed the Einstein and Planck gauges. \ In both gauges,
the extra or parallel force (per unit mass) of (5.15) leads to a
relation between the increments in the coordinates and momenta
(5.17) which is reminiscent of Heisenberg's relation. \ Further
study confirms that the extra force - which is inertial is the
Einstein sense of coming from the motion of the (extra part of
the) coordinate frame - results in Heisenberg's uncertainty
relation (5.20). \ A corollary of this is that the inertial rest
mass of a particle is quantized,
though the unit is set by the cosmological constant and is less than $%
10^{-65}$ gm by (5.21) and so too small to be observed using
current methods. \ These results prompted the brief study in
Section 6 of waves in $N(>4)D$. \ In 5D, what is in effect a Wick
rotation of the extra part of the null metric (6.1) was found to
lead to a wave function (6.3) which satisfies a hierachy of wave
equations (6.4). \ The first of these (6.5) is a restatement of
the standard prescription for the derivation of 4-momenta from a
wave function. \ The second (6.6) splits naturally into two
parts, of which one is a restatement of the geodesic equation of
classical theory (6.7), while the other is equivalent to the
Klein-Gordon equation of relativistic quantum theory (6.9). \ It
should be noted that the latter was derived without the use of
operators, and of course contains the Schrodinger equation in the
non-relativistic limit. \ More importantly, it should be noted
that solutions of the 5D field equations exist which have complex
metrics but which result (because of the structure of the field
equations) in measurable quantities which are real. \ An example
is (6.10), which is the canonical solution for a wave in a 5D
manifold. It describes an oscillation of the extra coordinate in
and out of the hypersurface called spacetime. \ The mass of the
particle associated with the wave also oscillates, with a mean
value which is zero, in accordance with inflationary quantum
field theory and the flatness of the 5D manifold. \ However, the
square of the mass is finite and in fact set by (6.11), which
says that the square of the inverse Compton wavelength of the
particle is proportional to the Ricci scalar of the embedded and
curved 4D manifold. \ This last property is manifestly Machian,
and is expected to be generic to $N(>4)D$ manifolds which contain
submanifolds that proscribe the 4D ``boxes'' which particles
inhabit. \ The general prescription for dynamics in\
N-dimensional spaces would appear to be a null product of
N-vectors (specifying the coordinate velocities in the classical
case or the wave numbers in the quantum case). \ This means that
particles are in causal contact in\ ND even though they appear to
be out of contact in 4D, so there are obvious implications for
the Aharanov-Bohm effect and the double-slit experiment. \ These
phenomena, and other consequences for membranes and strings
especially, will surely repay further investigation.

\bigskip

\section{\protect\underline{Acknowledgements}}

\qquad Thanks for academic comments go to A.D.\ Linde, H. Liu and B.
Mashhoon; and for financial support to the Natural Sciences and Engineering
Research Council of Canada. \ This work is dedicated to the memory of
Professor J. Moriaty, whose monograph \underline{The Dynamics of an Asteroid}
``ascends to such rarefied heights of pure mathematics that it is said that
there was no man in the scientific press capable of criticizing it'' [71].

\section{\protect\underline{References}}

\begin{enumerate}
\item  V. Fock, Zeit. Phys. (Leipzig) \underline{39}, 226 (1926).

\item  E. Leibowitz, N. Rosen, Gen. Rel. Grav. \underline{4}, 449 (1973).

\item  D. Kovacs, Gen. Rel. Grav. \underline{16}, 645 (1984).

\item  J. Gegenberg, G. Kunstatter, Phys. Lett. A \underline{106}, 410
(1984).

\item  A. Davidson, D.A. Owen, Phys. Lett. B \underline{177}, 77 (1986).

\item  J.A. Ferrari, Gen. Rel. Grav. \underline{21}, 683 (1989).

\item  B. Mashhoon, H. Liu, P.S. Wesson, Phys. Lett. B \underline{331}, 305
(1994).

\item  P.S. Wesson, J. Ponce de Leon, Astron. Astrophys. \underline{294}, 1
(1995).

\item  H. Liu, P.S. Wesson, Class. Quant. Grav. \underline{14}, 1651 (1997).

\item  B. Mashhoon, P.S. Wesson, H. Liu, Gen. Rel. Grav. \underline{30}, 555
(1998).

\item  P.S. Wesson, Space-Time-Matter (World Scientific, Singapore, 1999).

\item  W.N. Sajko, A.P. Billyard, Gen. Rel. Grav., in press (2001).

\item  D. Youm, hep-th/0004144 (2000).

\item  R. Maartens, hep-th/0004166 (2000).

\item  A. Chamblin, hep-th/0011128 (2000).

\item  A.P. Billyard, A.A. Coley, J.E. Lidsey, U.S. Nilsson, Phys. Rev. D%
\underline{61}, 043504 (2000).

\item  Y.M. Cho, D.H. Park, Gen. Rel. Grav. \underline{23}, 741 (1991).

\item  P.S. Wesson, B. Mashhoon, H. Liu, W.N.\ Sajko, Phys. Lett. B
\underline{456}, 34 (1999).

\item  P. West, Introduction to Supersymmetry and Supergravity\ (World
Scientific, Singapore, 1986).

\item  M.B. Green, J.H. Schwarz, E. Witten, Superstring Theory (Cambridge
Un. Press, Cambridge, 1987).

\item  S.S. Seahra, P.S. Wesson, Gen. Rel. Grav., in press (2001).

\item  A.D. Linde, Inflation and Quantum Cosmology (Academic Press, Boston,
1990).

\item  J.D. Bekenstein, Phys. Rev. D\underline{15}, 1458 (1977).

\item  P.S. Wesson, B. Mashhoon, H. Liu, Mod. Phys. Lett. A \underline{12},
2309 (1997).

\item  P.S. Wesson, H. Liu, Astrophys. J. \underline{440}, 1 (1995).

\item  A.P. Billyard, P.S. Wesson, Gen. Rel. Grav. \underline{28}, 129
(1996).

\item  J.E. Campbell, A Course of Differential Geometry (Clarendon Press,
Oxford, 1926).

\item  S. Rippl, C. Romero, R. Tavakol, Class. Quant. Grav. \underline{12},
2411 (1995).

\item  C. Romero, R. Tavakol, R. Zalaletdinov, Gen. Rel. Grav. \underline{28}%
, 365 (1996).

\item  J.E. Lidsey, C. Romero, R. Tavakol, S. Rippl, Class. Quant. Grav.
\underline{14}, 865 (1997).

\item  P.S. Wesson, J. Ponce de Leon, Astron. Astrophys. \underline{294}, 1
(1995).

\item  L.D. Landau, E.M. Lifshitz, The Classical Theory of Fields, p. 251
(Pergamon Press, Oxford, 1975).

\item  P.S. Wesson, H. Liu, Int. J. Th. Phys. \underline{36}, 1865 (1997).

\item  J. Ponce de Leon, Gen. Rel. Grav. \underline{20}, 539 (1988).

\item  P.S. Wesson, J. Ponce de Leon, J. Math. Phys. \underline{33}, 3883
(1992).

\item  J.L. Synge, A. Schild, Tensor Calculus (Dover, New York, 1978).

\item  R.G. Carlberg, Astrophys. J. \underline{375}, 429 (1991).

\item  M. Chiba, Y. Yoshii, Astrophys. J. \underline{510}, 42 (1999).

\item  J.M. Eppley, R.B. Partridge, Astrophys. J. \underline{538}, 489
(2000).

\item  C.M. Will, \underline{in} Gravitation: \ A Banff Summer Institute, p.
439 (eds. R.B. Mann, P.S. Wesson, World Scientific, Singapore, 1991).

\item  G. Leibbrandt, Non-Covariant Gauges (World Scientific, Singapore,
1994).

\item  W.N. Sajko, P.S. Wesson, H. Liu, J. Math. Phys. \underline{39}, 2193
(1998).

\item  S. Weinberg, Rev. Mod. Phys. \underline{61}, 1 (1989).

\item  Y.J. Ng, Int. J. Mod. Phys. D \underline{1}, 145 (1992).

\item  P.S. Wesson, Int. J. Mod. Phys. D \underline{6}, 643 (1997).

\item  P.W. Bridgman, Dimensional Analysis (Yale Un. Press, New Haven, 1922).

\item  E.A. Desloge, Am. J. Phys. \underline{52}, 312 (1984).

\item  P.S. Wesson, Sp. Sci. Rev. \underline{59}, 365 (1992).

\item  W.N. Sajko, Phys. Rev. D \underline{60}, 104038 (1999).

\item  W.N. Sajko, Int. J. Mod. Phys. D \underline{9}, 445 (2000).

\item  W.N. Sajko, P.S. Wesson, Gen. Rel. Grav. \underline{32}, 1381 (2000).

\item  J.B. Hartle, S.W. Hawking, Phys. Rev. D \underline{28}, 2960 (1983).

\item  A. Vilenkin, Phys. Lett. B \underline{117}, 25 (1982).

\item  F. Darabi, W.N. Sajko, P.S.\ Wesson, Class.\ Quant. Grav. \underline{%
17}, 4357 (2000).

\item  D. Kramer, H. Stephani, E. Herlt, M. MacCallum, E.\ Schmutzer, Exact
Solutions of Einstein's Field Equations (Cambridge Un. Press, Cambridge,
1980).

\item  T. Appelquist, A. Chodos, Phys. Rev. D \underline{28}, 772 (1983).

\item  H. Liu, S. Wanzhong, Gen. Rel. Grav. \underline{20}, 407 (1988).

\item  H. Liu, P.S.\ Wesson, Int. J. Mod. Phys. D \underline{7}, 737 (1998).

\item  H. Liu, P.S. Wesson, Mod. Phys. Lett. A \underline{13}, 2689 (1998).

\item  W.N. Sajko, P.S. Wesson, H. Liu, J. Math. Phys. \underline{40}, 2364
(1999).

\item  E.T. Newman, A.J. Janis, J. Math. Phys. \underline{6}, 915 (1965).

\item  E.T. Newman, E. Couch, K. Chinnapared, A. Exton, A. Prakash, R.\
Torrence, J.\ Math. Phys. \underline{6}, 918 (1965).

\item  E.J. Flaherty, \underline{in} General Relativity and Gravitation, p.
207 (ed. A. Held, Plenum Press, New York, 1980).

\item  C.P. Boyer, J.D. Finley, J.F. Plebanski, \underline{in} General
Relativity and Gravitation, p. 241 (ed. A.\ Held, Plenum Press, New York,
1980).

\item  S. Chakraborty, P. Peldan, Int. J. Mod. Phys. D \underline{3}, 695
(1994).

\item  S.P. Drake, P. Szekeres, gr-qc/9807001 (1998).

\item  K. Lake, P. Musgrave, D. Pollney, GR Tensor Version 1.19 (Queen's
Un., Kingston, 1995).

\item  A. Macias, O. Obregon, G.J. Fuentes y Martinez, Gen. Rel. Grav.
\underline{25}, 549 (1993).

\item  A. Macias, H. Dehnen, \underline{in} Proc. 5th Canadian Conference on
General Relativity and Relativistic Astrophysics, p. 443 (eds. R.B. Mann,
R.G.\ McLenaghan, World Scientific, Singapore, 1994).

\item  P. Wesson, W.N. Sajko, \underline{in} Proc. 8th Canadian Conference
on General Relativity and Relativistic Astrophysics, p. 262 (eds. C.P.
Burgess, R.C.\ Myers, World Scientific, Singapore, 1999).

\item  A.C.\ Doyle, \underline{in} The Complete Sherlock Holmes Long
Stories, p. 409 (Murray, London, 1929). \ See also \underline{in} The
Complete Sherlock Holmes Short Stories, p. 580 (Murray, London, 1928).
\end{enumerate}

\end{document}